\documentstyle[11pt,paspconf,epsf]{article}

\begin{document}

\title{Cluster Distances: the Good, the Bad and the Ugly}
\author{John R. Lucey$^1$, Russell J. Smith$^1$, Michael J. Hudson$^2$,\\
David J. Schlegel$^3$ and Roger L. Davies$^1$
}
\affil{$^1$\,Department of Physics, University of Durham,
Science Laboratories, South Road, Durham, DH1 3LE, United Kingdom.}
\affil{$^2$\,Department of Physics \& Astronomy, University of Victoria, \\
PO Box 3055, Victoria BC V8W 3PN, Canada.}
\affil{$^3$\,Department of Astrophysical Sciences, Princeton University, \\
Peyton Hall, Princeton NJ, USA.}

\begin{abstract}
Galaxy clusters are important targets for peculiar 
velocity studies as a direct comparison of the various 
distance indicators can be made. 
The potential
problem of an environmental effect biasing the
distances to clusters of different richness is examined. 
The level of intrinsic cluster-to-cluster 
variations is also reviewed. 
We investigate the claim that
the derived Fundamental Plane (FP)
distances vary systematically with FP scatter.
The cluster distances and peculiar
velocities derived from
Tully-Fisher, FP and Brightest Cluster Galaxies 
are compared and we find good agreement.
\end{abstract}

\section{Introduction}

A number of early attempts to measure cluster distances 
could be described as {\it ugly} by current standards. 
In particular, many early studies
suffered from poorly determined systematic errors.
A few early studies were wrong due to 
erroneous data, i.e. {\it bad}. As the field
has developed a considerably better understanding of 
the practical
problems associated with cluster distance measurement have
emerged and hopefully reliable ({\it good}) distances are
now starting to appear. The new surveys are starting to
provide a sizable set of clusters that possess both good 
Tully-Fisher (TF) and Fundamental Plane (FP) data.

Here we address three aspects of cluster distances:
\begin{itemize}
\item At what level do environmentally dependent 
stellar population biases affect the
FP cluster distance determinations?
\item Do the derived FP distances vary systematically with FP scatter? 
\item How well do the different cluster distance indicators compare?
\end{itemize}

We use the SMAC FP dataset to investigate these themes. 
For this dataset the random and systematic 
errors in the velocity dispersion ($\sigma$) 
measurements are 0.025 dex and 0.015 dex respectively.
For the $Mg_2$ indices these errors are
0.010 mag and 0.002 mag respectively.
Hence this dataset is the highest precision and largest FP dataset that 
is currently available.

\section{Stellar population biases in the FP cluster distances}

The systematic variation in the stellar populations of 
elliptical and S0 galaxies with the environment has 
long been suggested, e.g. Larson, Tinsley \& Caldwell (1980).
While such an effect on the FP distance indicator is probably
only at the level of $\sim$0.02 dex between the dense cluster 
cores and isolated field regions (Lucey 1995), this may produce
a bias in the derived distances for poor and
rich clusters. 

We have examined the cluster-to-cluster offsets in
the $Mg_2-\sigma$ relation.
Following Colless et al. (1999), in Figure 1 we show
the median $\Delta Mg_2$ and uncertainty for each cluster. 
The SMAC clusters display a remarkable degree of homogeneity
with a measured scatter in $\Delta Mg_2$ of only 0.006 mag. 
If the cluster $Mg_2-\sigma$ relation is universal 
then there is a probability of 0.045 that the measured 
$\chi^2$ would be observed. In order to derive a 
reduced $\chi^2$ of unity an intrinsic rms scatter of 
0.004 mag must be added. As the expected 
$\Delta Mg_2$ variation due to the known level of systematic errors 
is $\sim$0.003 mag, there is no evidence for any significant 
cluster-to-cluster variation in the
$Mg_2-\sigma$ relations. The true intrinsic 
$\Delta Mg_2$ scatter is less than 0.004 mag. 
This limit is 
4$\times$ smaller than that found for the 
EFAR sample (Colless et al. 1999).

\begin{figure}
\plotone{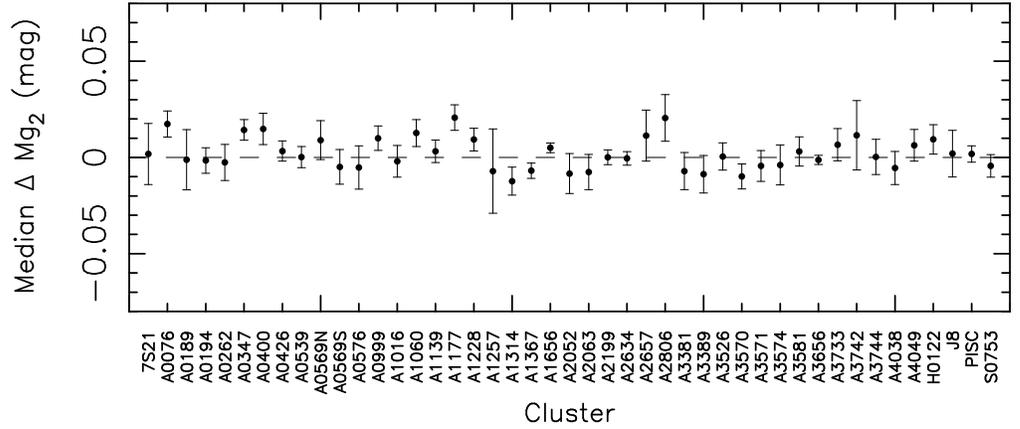}
\caption{Variation of the cluster
$Mg_2-\sigma$ relation zero-points.
The median $\Delta Mg_2$ values and their uncertainties are displayed 
for the 45 SMAC clusters that have $Mg_2$ measurements.  Three clusters 
(i.e. A76, A347 and A1177) are offset by more than two standard deviations.}
\label{fig1}
\end{figure}

Stellar population models imply that if $Mg_2$ differences are
caused by age differences then a
0.004 mag change in $\Delta Mg_2$ translates into a
luminosity change that would give an apparent distance
shift of 0.02 dex (see e.g. Jorgensen et al. 1996).
This is equivalent to a 340 km\,s$^{-1}$ shift
at the Coma cluster distance.
If stellar population differences were 
responsible for some component of the 
FP peculiar velocities we would
predict these to be correlated with $\Delta Mg_2$.
Following Burstein, Faber \& Dressler (1990), 
we use the quantity $\log_{10}$(V$_{CMB}$/R$_{CL})$ 
as a measure of the cluster peculiar velocity.
V$_{CMB}$ is the cluster velocity in the local CMB 
rest frame and R$_{CL}$ is the derived cluster distance. 
In Figure 2, we present the correlation between
$\Delta Mg_2$ and $\log_{10}$(V$_{CMB}$/R$_{CL})$. 
If the galaxy luminosities are affected by 
age differences then a slope of --5 is expected.
Unfortunately the relatively large errors 
do not allow a meaningful constraint to be 
placed on this correlation;
the best fit slope allowing for the errors in both
co-ordinates is --18\,$\pm$90.
For clusters with $\Delta Mg_2$\,$<$\, 0.0 the mean
$\log_{10}$(V$_{CMB}$/R$_{CL})$ value is 0.011\,$\pm$\,0.008
whereas for clusters with $\Delta Mg_2$\,$>$\, 0.0 the value is
0.004\,$\pm$\,0.008. We conclude that there is no evidence that
age differences bias the FP distances for the SMAC sample.

\begin{figure}
\plottwo{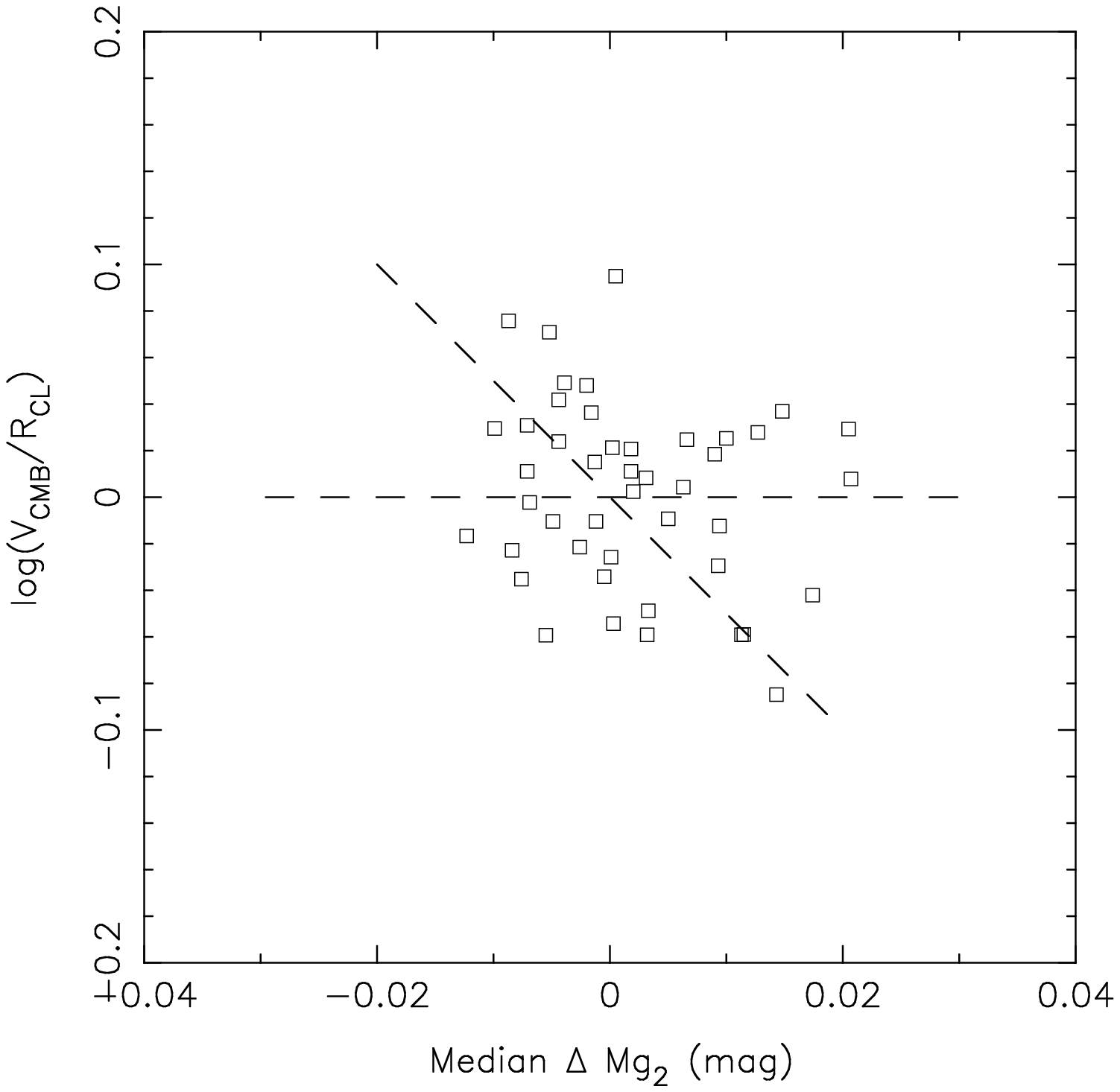}{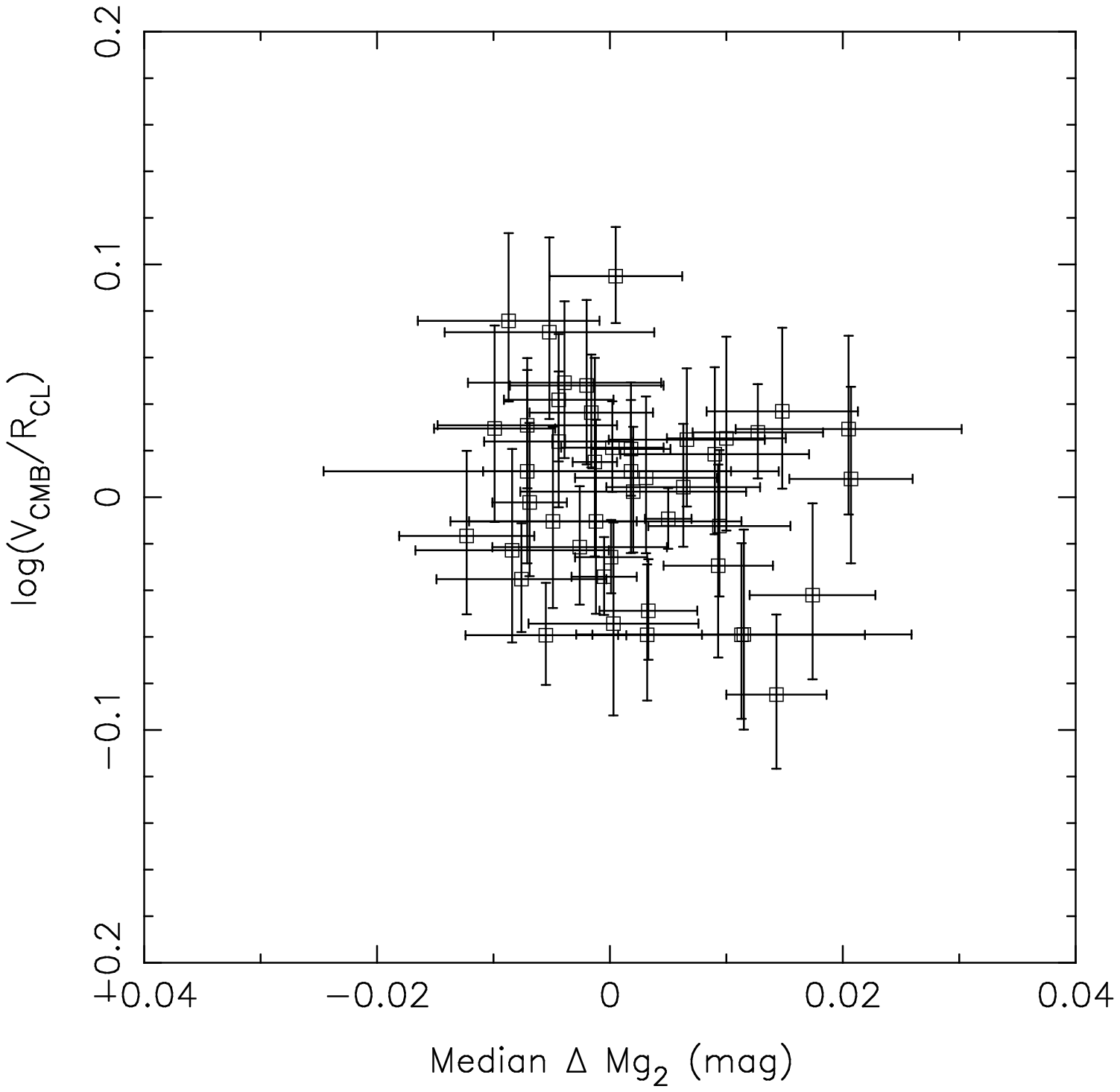}
\caption{Is cluster peculiar velocity correlated with $\Delta Mg_2$\,?
The median $\Delta Mg_2$ for each SMAC cluster is plotted
against $\log_{10}$(V$_{CMB}$/R$_{CL})$. The left hand
panel shows the data with no error bars displayed. The
inclined dashed line shows the expected correlation if
the galaxy luminosities are affected by small age
differences. If all the cluster galaxies in the
sample were homogeneous then the data is expected to display
no correlation, i.e. the horizontal dashed line.
The right hand panel shows the data with the error bars plotted.} 
\label{fig2}
\end{figure}

If such a stellar population bias existed, in order
to cause a spurious bulk flow signal the $\Delta\,Mg_2$ residuals
would need to be correlated on the sky. For the SMAC
sample this is not the case and the bulk flow 
measurement is only weakly affected. If a $\Delta Mg_2$ term 
is included in the distance indicator the SMAC bulk 
flow measurement {\it increases} by $\sim$70 km\,s$^{-1}$.

\section{Is the cluster peculiar velocity related to FP scatter?}

Using mostly literature FP data for 20 rich clusters,  
Gibbons, Fruchter and Bothun (1998) have recently 
suggested that clusters which have a lower FP rms scatter
also possess a lower peculiar velocity and vice versa.
In the left hand panel of Figure 3 we reproduce their plot 
of FP scatter versus peculiar velocity. They propose two sub-groups
with the division at an FP rms scatter of $\sim$0.07 in units of 
$\log_{10}\sigma$. For this sample, the lower 
FP scatter clusters have an average absolute 
peculiar velocity of 220\,$\pm$\,50 km\,s$^{-1}$ (n=11) 
whereas the higher FP scatter clusters have a value of
830\,$\pm$\,200 km\,s$^{-1}$ (n=9). They suggest 
that the measured
peculiar velocities for the higher FP scatter group of 
clusters are not real but due to a systematic
difference in the FP relation.

The equivalent plot for the SMAC clusters is shown
in the right hand panel of Figure 3. 
For the 39 clusters which have an rms FP scatter
less than 0.07, the average 
absolute peculiar velocity is 610\,$\pm$\,90 km\,s$^{-1}$.
For the 17 clusters with an FP rms scatter greater than 0.07,
the average value is 760\,$\pm$\,120 km\,s$^{-1}$.
If we consider only the well sampled clusters 
($n_{GAL}\,\geq$\,10), which are less affected by 
outliers, the lower and higher FP scatter groups have
an average absolute peculiar velocity of
560\,$\pm$\,110 km\,s$^{-1}$ (n=16) and
690\,$\pm$\,150 km\,s$^{-1}$ (n=9) respectively.
Thus for the SMAC sample there is no evidence
for a trend of peculiar velocity with FP scatter. 

\begin{figure}
\plottwo{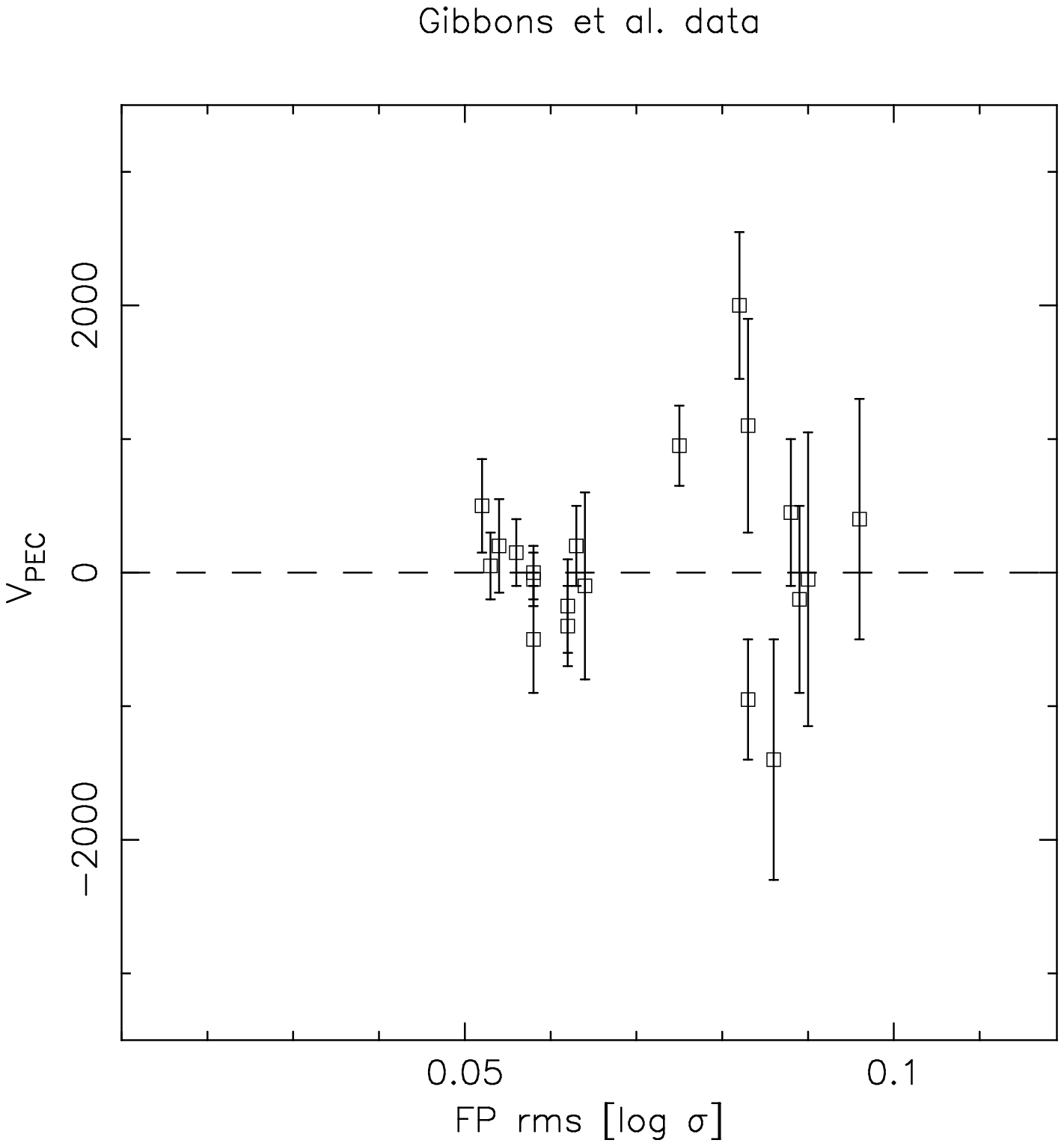}{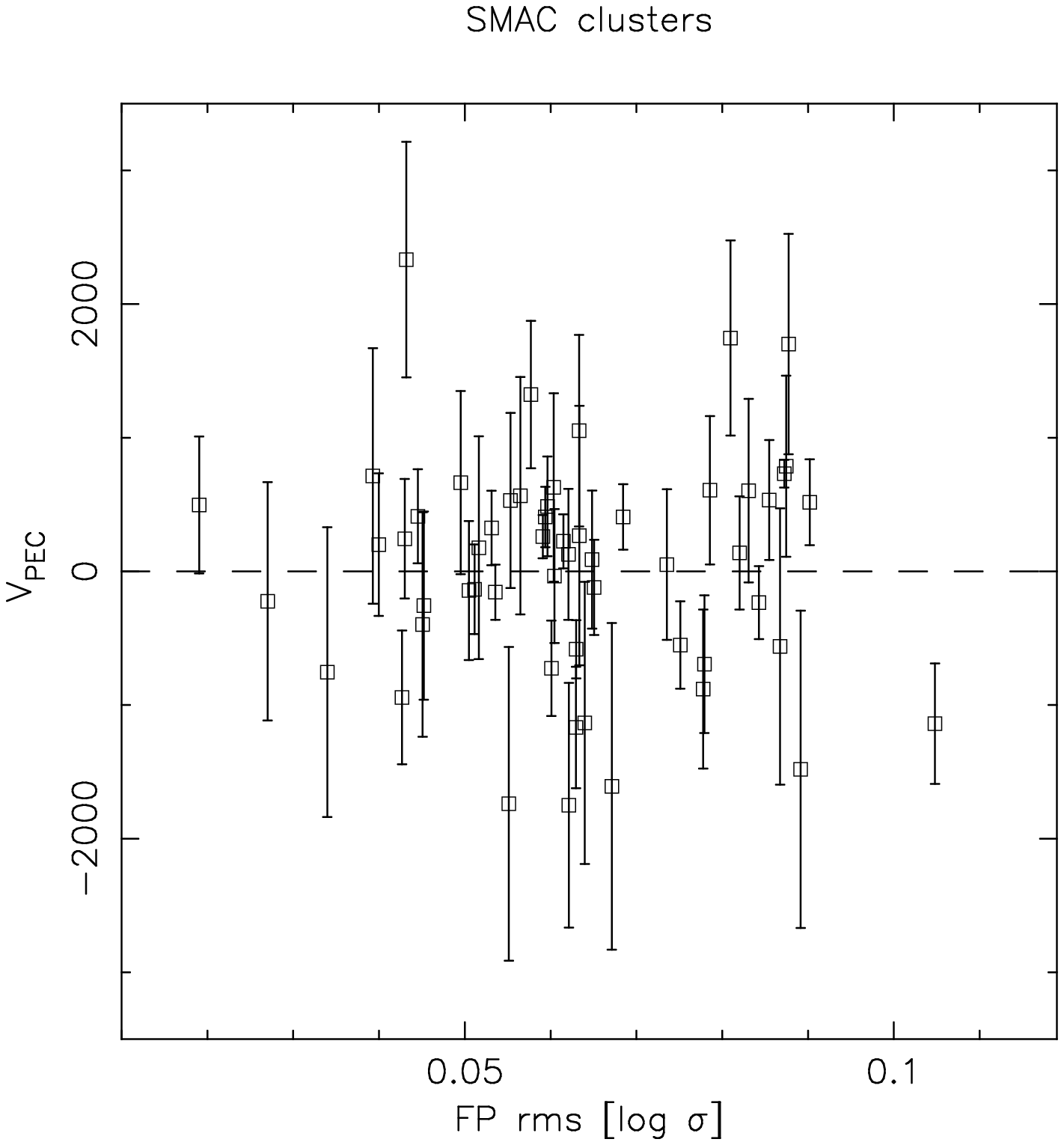}
\caption{The relationship between FP scatter and peculiar velocity.
The left hand panel displays the Gibbons et al. data and the right hand
panel displays the SMAC cluster data. 
} 
\label{fig3}
\end{figure}

\section{Comparison of the BCG, TF and FP results}

Direct comparison of the cluster distances derived from
different indicators is a valuable test of
systematics. In this section
we compare results from the FP work of SMAC with
the Brightest Cluster Galaxy (BCG) 
study of Lauer \& Postman (1984) (LP) and
the TF study of Giovanelli et al. (1999) (SCI/II). 
The average errors in the cluster distance measurements
for the SMAC, LP and SCI/II samples are 0.026, 0.081 
and 0.022 dex respectively.

One potential problem is that different
surveys don't always select
precisely the same physical structure. In particular, FP samples 
concentrate on the cluster core whereas
TF samples cover a wider spatial area. 
If different physical structures are selected
then this may result in a disagreement 
in the mean redshift adopted for the cluster. 
In Figure 4 we compare the cluster redshifts used 
by SMAC with those of LP and SCI/II. 
The agreement is good with an average absolute velocity 
difference of 180 km\,s$^{-1}$.  
\begin{figure}
\plottwo{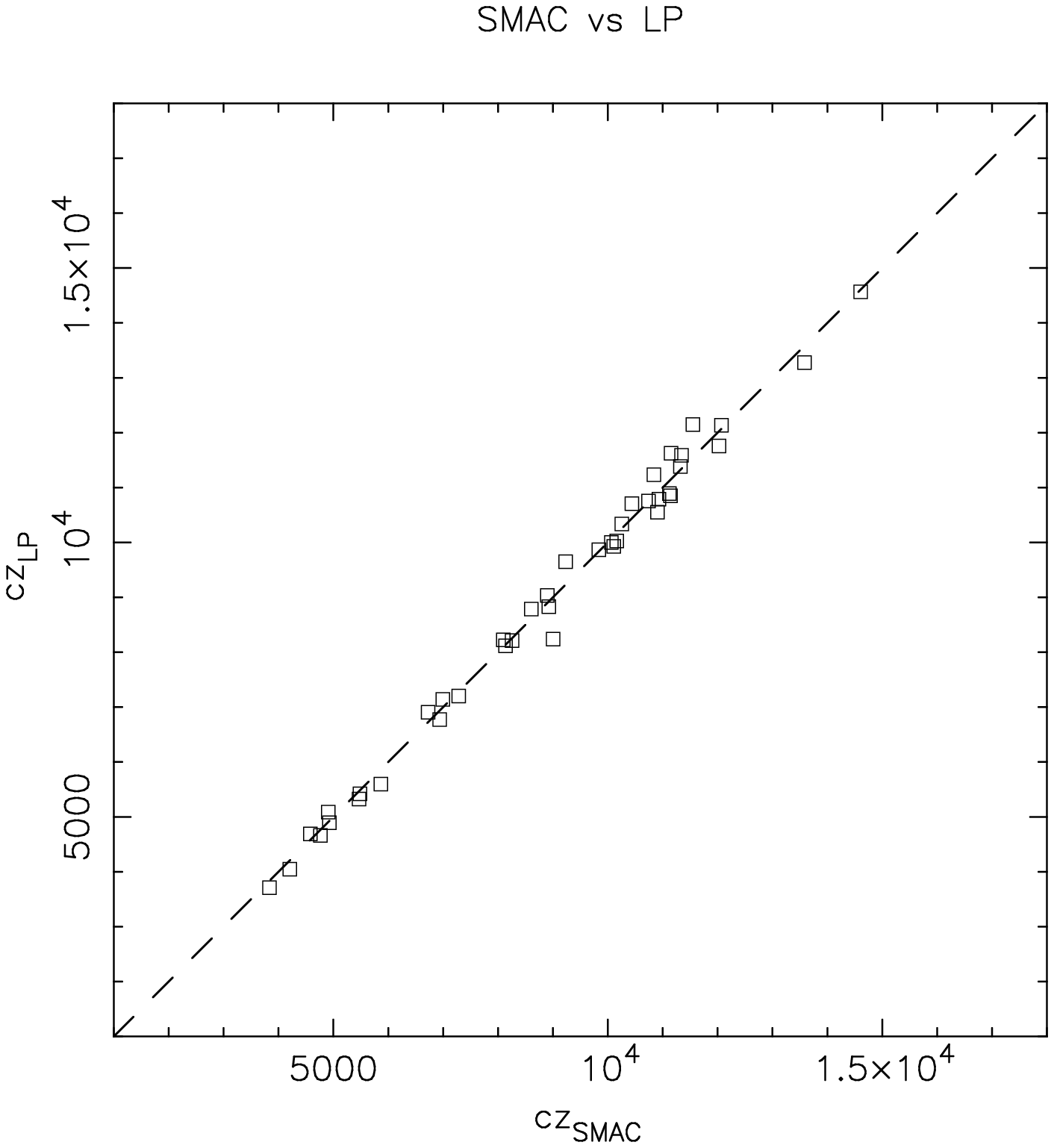}{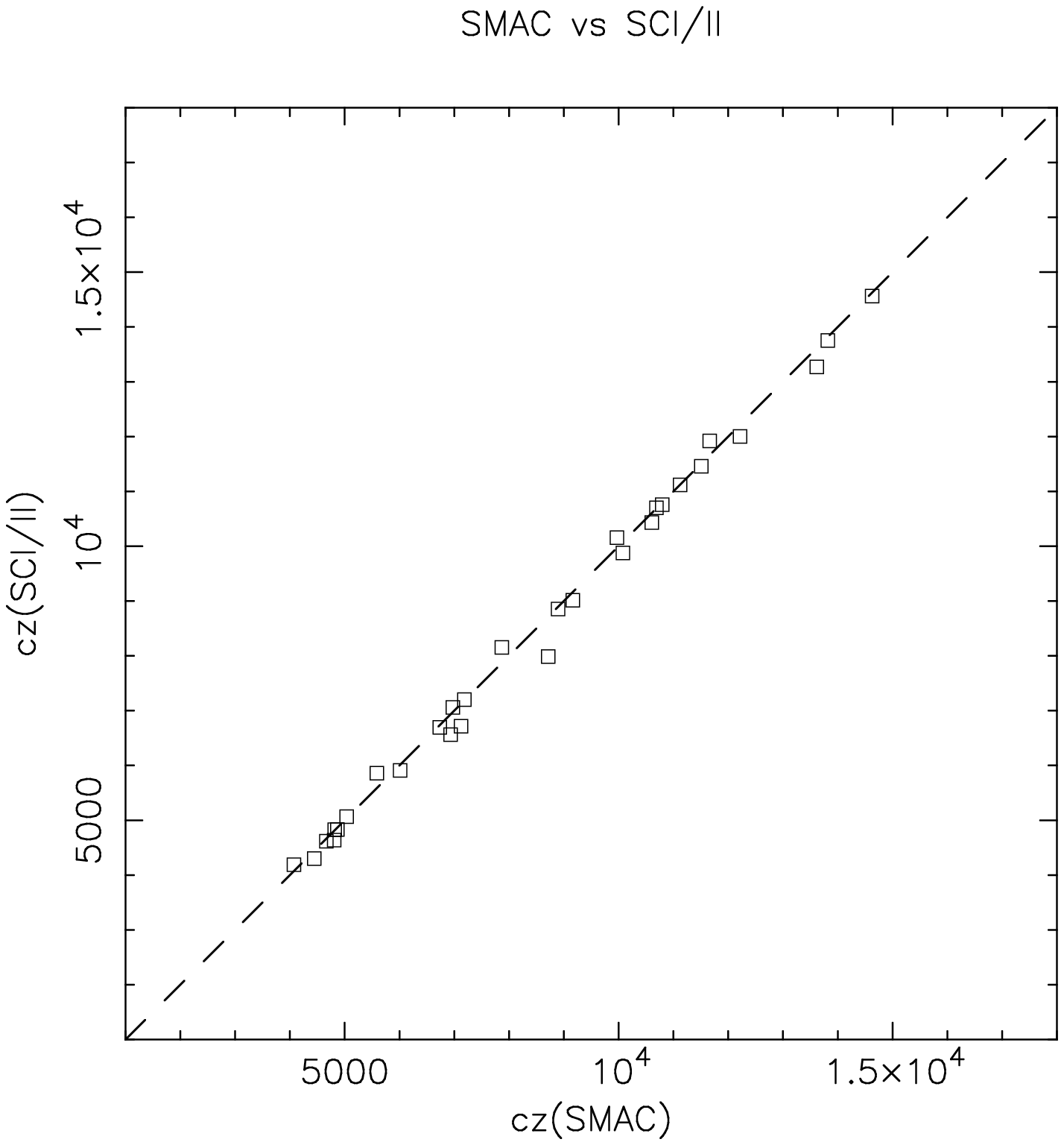}
\caption{Comparison of the SMAC cluster velocities with those of
LP and SCI/II. The most 
discrepant cluster is A4038 for which both 
LP and SCI/II have a velocity of 740 km\,s$^{-1}$ 
less than that used by SMAC. This difference is
due to confusion with the adjacent A4049 cluster.}
\label{fig4}
\end{figure}

In Figure 5 we present the distance-distance comparisons of
SMAC with LP and SCI/II. The SMAC and LP distances
are in reasonable agreement; there is a  
probability of 0.06 that the measured $\chi^2$ 
would arise given the distance errors. 
The SMAC and SCI/II distances are in better
agreement with a $\chi^2$ probability 
of 0.13. 

\begin{figure}
\plottwo{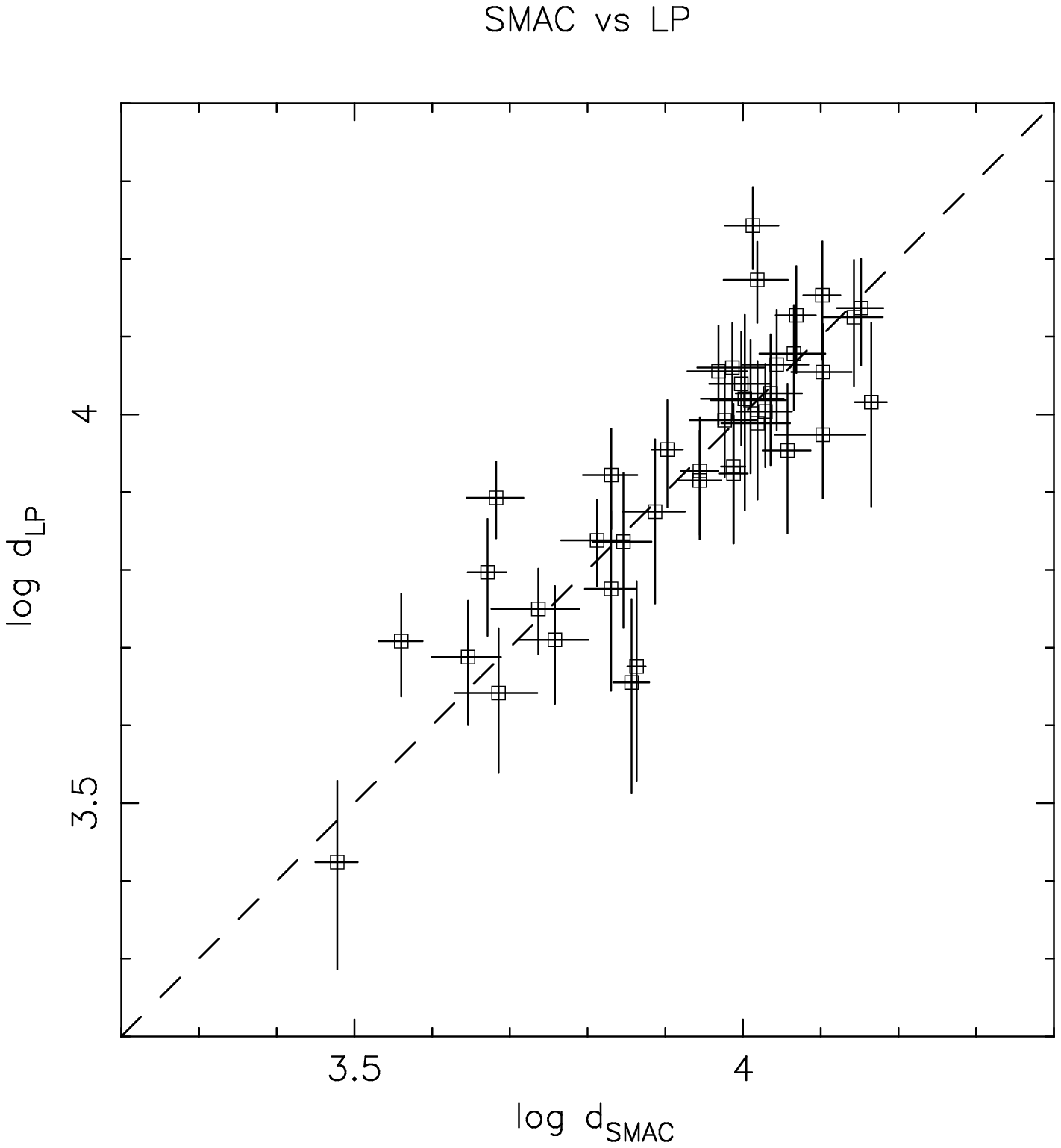}{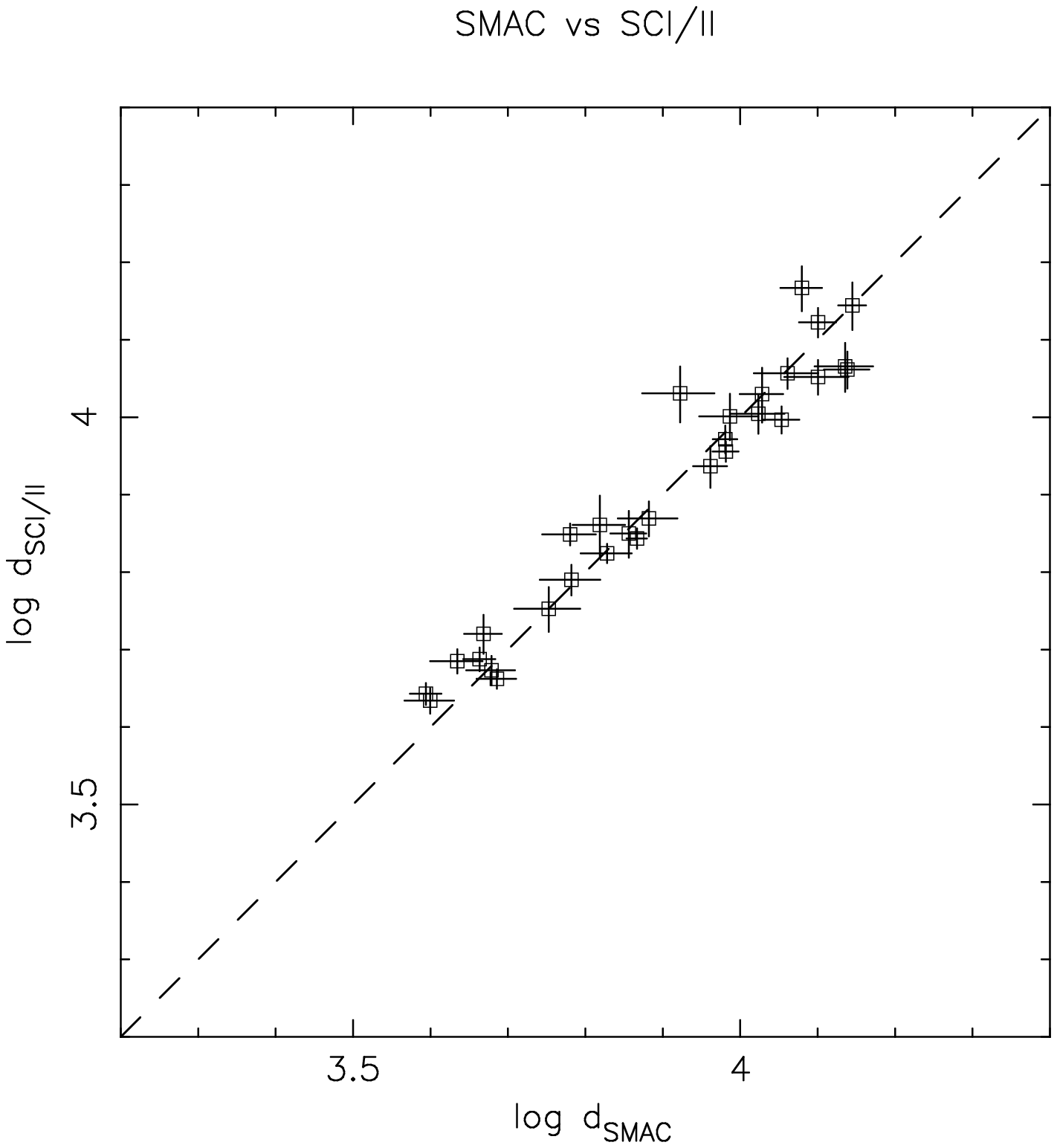}
\caption{Comparison of the SMAC cluster distances with those of
LP and SCI/II. For the SMAC-LP comparison there are 41 clusters in
common. Four clusters are formally discrepant; 
A262 (3.4$\sigma$),
Hydra (=A1060) (2.1$\sigma$),
A3381 (3.7$\sigma$),
and A3733 (2.3$\sigma$).
For the SMAC-SCI/II comparison there are 30 clusters in
common; Centaurus (A3526) is excluded from the comparison because
of the different treatment of the Cen30 and Cen45 subcomponents.
Two clusters are discrepant;
A957 (2.2$\sigma$) and Hydra (2.0$\sigma$).
}
\label{fig5}
\end{figure}

\begin{figure}
\plottwo{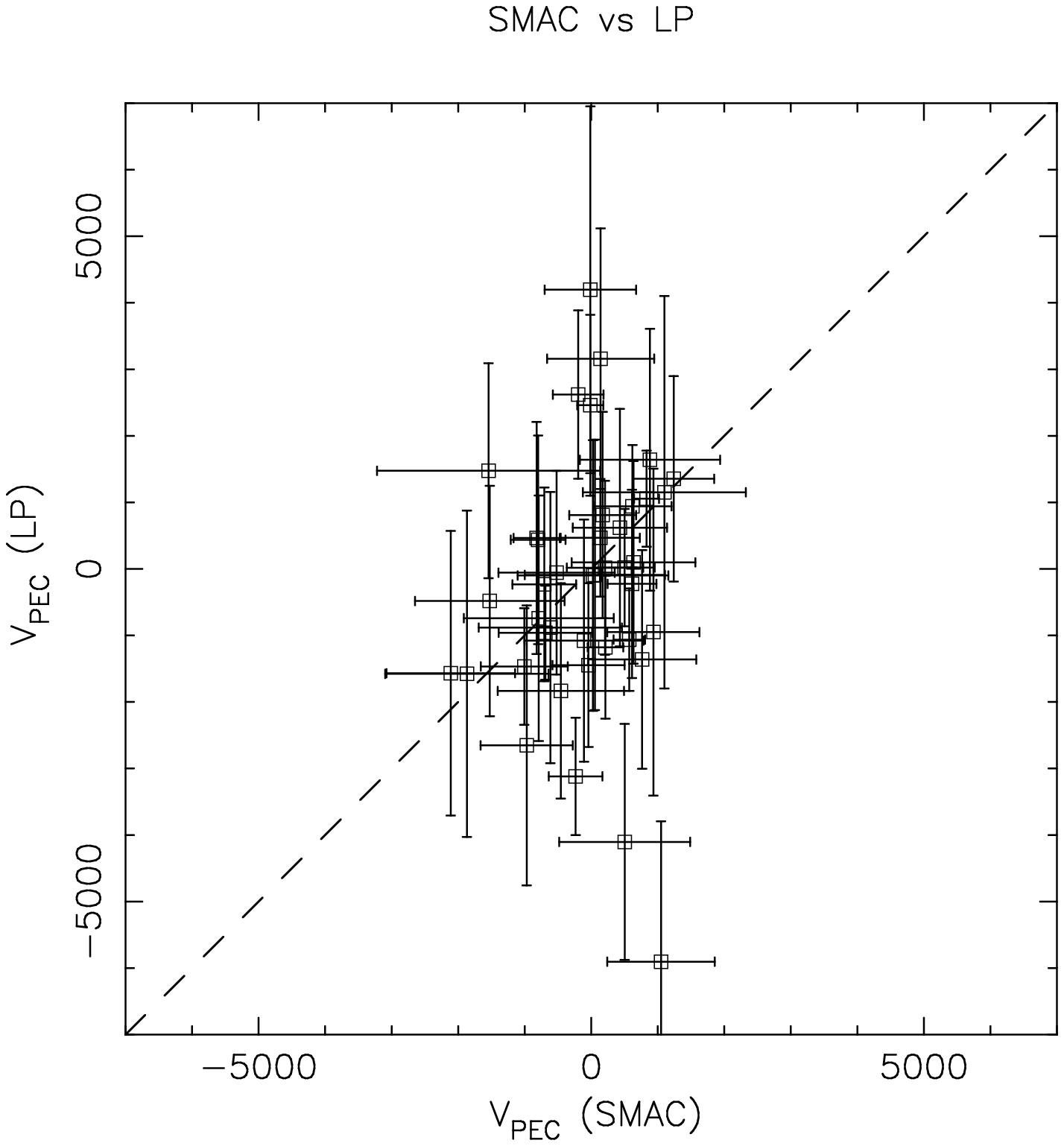}{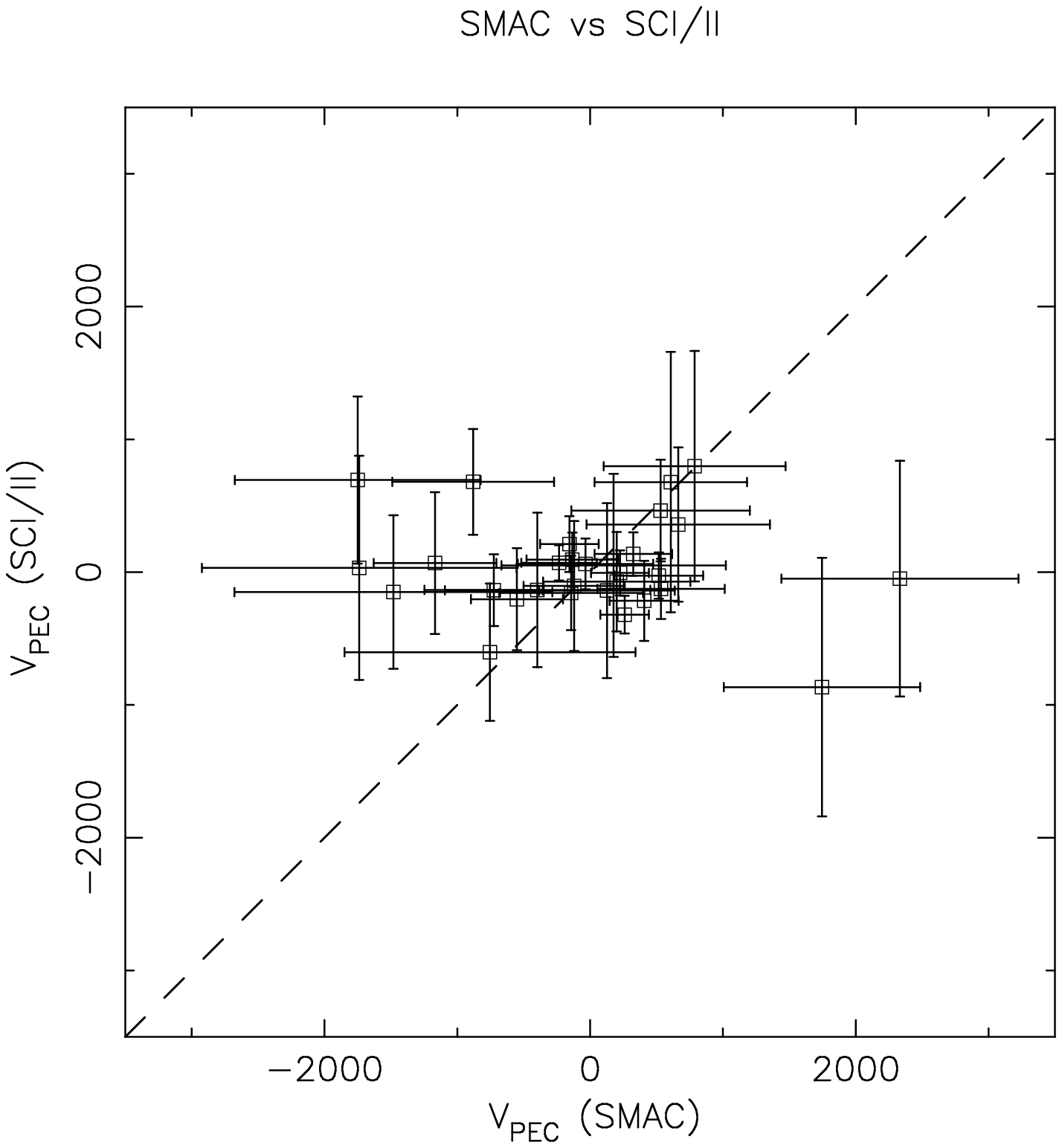}
\caption{Comparison of the SMAC peculiar velocities
with those of LP and SCI/II. The scale for the right hand
panel is half that of the left hand panel.
For the SMAC-LP comparison, the larger error for LP 
distance measurements spreads the clusters vertically.
For the SMAC-SCI/II comparison a few discrepant 
clusters are spread horizontally becauses SMAC measures, 
on average, larger individual cluster peculiar velocities.
}
\label{fig6}
\end{figure}

The SMAC-LP and SMAC-SCI/II
peculiar velocity comparisons are presented in Figure 6. 
For both comparisons the agreement is again reasonable for 
the measurement errors. For the SMAC-LP
comparsion there is a probability of 
0.15 that the observed $\chi^2$ would arise given the 
peculiar velocity errors. For the SMAC-SCI/II comparison 
the probability is 0.08. In this comparison most clusters
are in good agreement and form a tight group of points
at the centre of the plot. This is further illustrated
in Figure 7 where we display the differences in the
measured peculiar velocities.
The most discrepant cluster is
Hydra(A1060); $V_{PEC}$(SMAC)\,=\,+260\,$\pm$\,183 km\,s$^{-1}$,
$V_{PEC}$(SCI/II)\,=\,--320\,$\pm$\,142 km\,s$^{-1}$.

\begin{figure}
\plotone{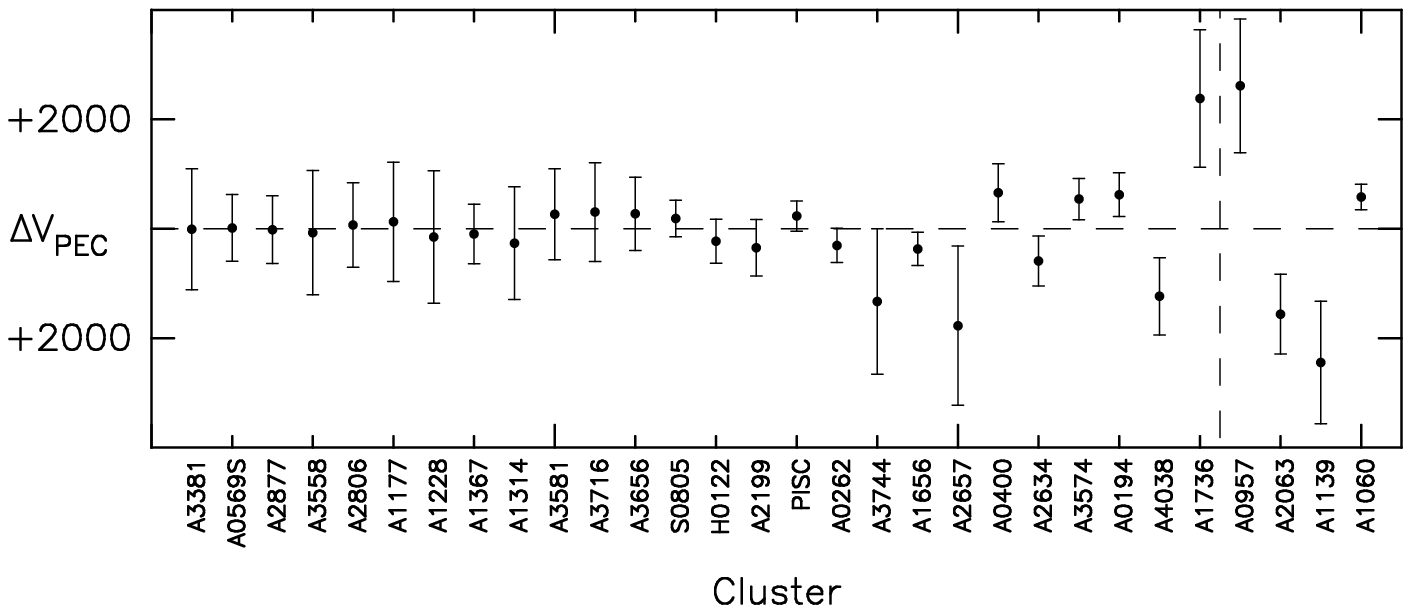}
\caption{The differences in the cluster peculiar velocities as
measured by SMAC and SCI/II.
The clusters are plotted in order of the significance of 
the difference. Clusters that lie to the right of the
vertical dashed line are discrepant at greater than the 
2$\sigma$ level.
}
\label{fig7}
\end{figure}

\section{Conclusions}

We have used the SMAC data to address
three aspects of cluster distances and have concluded that:

\begin{enumerate}
\item
{
While there is no evidence for significant differences
in the stellar populations of the SMAC FP cluster sample as 
determined by the $Mg_2-\sigma$ relation, this is a
relatively poor tool to assess if intrinsic cluster-to-cluster
variations bias the FP distances. 
}
\item
{
The SMAC dataset of 56 clusters does not support 
the claim made by Gibbons, Fruchter and Bothun (1998) that 
the FP scatter is related to cluster peculiar velocity.
}
\item
{
The SMAC FP and LP BCG distances are in
reasonable agreement with only four out of 41 clusters
being discrepant at greater than the 2$\sigma$ level.
The SMAC FP and SCI/II TF distances are in good agreement
with only two out of 30 clusters discrepant at greater than
the 2$\sigma$ level. 
}
\end{enumerate}


\begin{references}
\reference Burstein D., Faber S.M., Dressler A. 1990, \apj, 354, 18
\reference Colless M., Burstein D., Davies R.L.,
McMahan R.K., Saglia R.P, Wegner G. 1999, astro-ph/9811089
\reference Gibbons R.A., Fruchter A.S., Bothun G.D. 1998, astro-ph/9903380
\reference Giovanelli R., Dale D., Haynes M., Hardy E.,
Campusano L. 1999, astro-ph/9906362
\reference Jorgensen I., Franx M., Kjaergaard P. 1996, \mnras, 280, 167
\reference Larson R.B., Tinsley B.M., Caldwell C.N. 1980, \apj, 237, 692
\reference Lauer T. R., Postman M. 1994, \apj, 425, 418 
\reference Lucey J.R. 1995, {\it In}~: Van der Kruit P. \& Gilmore G. (Eds),
      {\it IAU Symposium 164: Stellar populations}, Kluwer Academic,
            p281
\end{references}
\end{document}